\input harvmac
\def\lf{16\pi^2}

\def\frak#1#2{{\textstyle{{#1}\over{#2}}}}
\def\frakk#1#2{{{#1}\over{#2}}}
\def\pa{\partial}
\def\semi{;\hfil\break}
\def\ga{\gamma}

\def\Xtilde{\tilde X}
\def\sy{supersymmetry}
\def\sic{supersymmetric}
\def\NSVZ{{\rm NSVZ}}

\def\DREDp{{\rm DRED}'}
\def\drbar{{\overline{\rm DR}'}}

\def\npb{{Nucl.\ Phys.\ }{\bf B}}

\def\prd{{Phys.\ Rev.\ }{\bf D}}

\def\plb{{Phys.\ Lett.\ }{\bf B}}

\def\zpc{Z.\ Phys.\ {\bf C}}

\def\ph{\phi}

\def\bM{M^*}

\def\tY{\tilde Y}

\def\lf{16\pi^2}

\def\Qcal{{\cal Q}}
\def\Rcal{{\cal R}}  
\def\Ocal{{\cal O}}
{\nopagenumbers
\line{\hfil LTH 454}
\line{\hfil hep-ph/9907255}
\line{\hfil Revised Version}
\vskip .5in
\centerline{\titlefont RG invariant solutions for the} 
\centerline{\titlefont soft supersymmetry breaking parameters}
\vskip 1in
\centerline{\bf I.~Jack and D.R.T.~Jones}
\bigskip
\centerline{\it Dept. of Mathematical Sciences,
University of Liverpool, Liverpool L69 3BX, U.K.}
\vskip .3in
We show that the results for  the soft couplings in supersymmetric gauge
theories recently derived via the  superconformal anomaly are equivalent
to results obtained earlier from a consideration of exact RG-invariant
trajectories. We give a full proof of these results for a general
multi-coupling theory. Moreover, we demonstrate the existence of a
distinct finite  RG-invariant solution for the soft couplings in the
case when the unbroken supersymmetric theory is finite. 
\Date{July 1999}}
\newsec{Introduction}
There has recently been interest in a specific and  predictive 
source of \sy-breaking originating in a vacuum
expectation value for an auxiliary field in the supergravity  multiplet,
rather than from a 
hidden sector~\ref\con{L. Randall and R. Sundrum, hep-th/9810155\semi
G.F. Giudice, M.A. Luty, H. Murayama and  R. Rattazzi,
JHEP 9812 (1998) 27\semi 
A. Pomarol and  R. Rattazzi, hep-ph/9903448\semi
T. Gherghetta, G.F. Giudice, J.D. Wells, hep-ph/9904378\semi
Z. Chacko, M.A. Luty, I. Maksymyk and E. Ponton, hep-ph/9905390\semi
E. Katz, Y. Shadmi and Y. Shirman, hep-ph/9906296}. 
The result is that the  various soft
breaking masses and interactions  (the gaugino mass $M$, the $\phi^3$
coupling $h^{ijk}$ and the  $\phi\phi^*$-mass $m^i_j$) are all given, via 
the superconformal anomaly, in
terms  of the gravitino mass and the $\beta$-functions of the unbroken
theory  by simple relations that are renormalisation group invariant  to
all orders in perturbation theory.  
On the other
hand, prior to this\ref\jjpa{I.~Jack, D.R.T.~Jones and A.~Pickering,
\plb426 (1998) 73}\ref\akk{L.V.~Avdeev, D.I.~Kazakov and
I.N.~Kondrashuk, \npb510 (1998) 289},  
it was shown how to write down RG-invariant
trajectories for the soft couplings, given RG-invariant trajectories for
the Yukawa couplings. In  this paper we shall show that 
this approach leads in fact to precisely the same results for the 
soft terms as the superconformal anomaly one, except that because 
there is no high-scale physics input, the gravitino mass is replaced 
by a constant of integration. 
We shall give a careful analysis demonstrating the RG-invariance of the 
results in the general multi-coupling case.
We also demonstrate the existence of a distinct RG-invariant 
solution for the case when the unbroken \sic\ theory is finite; 
corresponding to a class of all-orders finite theories 
with broken \sy\ (this is not, of course, a class of conformal 
non-\sic\ theories, because \sy\ is broken by dimensionful interactions).  

For a $N=1$ supersymmetric gauge theory with
superpotential
\eqn\eqf{W(\Phi)={1\over6}Y^{ijk}\Phi_i\Phi_j\Phi_k+   
{1\over2}\mu^{ij}\Phi_i\Phi_j,}
we take the soft breaking Lagrangian $L_{SB}$ as follows:
\eqn\Aaf{
L_{\rm SB}=(m^2)^j_i\ph^{i}\ph_j+
\left({1\over6}h^{ijk}\ph_i\ph_j\ph_k+{1\over2}b^{ij}\ph_i\ph_j
+ {1\over2}M\lambda\lambda+{\rm h.c.}\right)}
We assume that there are no gauge-singlet
fields and (for the time being) that $\cal G$ is simple.

In Refs.~\jjpa%
\nref\jj{I.~Jack and D.R.T.~Jones, \plb 415 (1997) 383}%
--\ref\jjpb{I.~Jack, D.R.T.~Jones and A.~Pickering, \plb432 (1998) 114}\
(see also Ref.~\ref\ahetal{N.~Arkani-Hamed et al, 
\prd58 (1998) 115005}) it was shown that the 
soft $\beta$-functions are given in terms of $\beta_g$
and the chiral supermultiplet anomalous dimension $\gamma^i{}_j$ as follows:
\eqna\betasoft$$\eqalignno{
\beta_M &= 2{\cal O}\left({\beta_g\over g}\right),
&\betasoft a\cr
\beta_h^{ijk} &= \gamma^{(i}{}_lh^{jk)l}-2\gamma_1^{(i}{}_lY^{jk)l},
&\betasoft b\cr
\beta_b^{ij} &= \gamma^{(i}{}_lb^{j)l}-2\gamma_1^{(i}{}_l\mu^{j)l},
&\betasoft c\cr
(\beta_{m^2})^i{}_j &= \left(2{\cal O}{\cal O}^* +2M\bM g^2{\partial       
\over{\partial g^2}} +\tY_{lmn}{\partial\over{\partial Y_{lmn}}}
+\tY^{lmn}{\partial\over{\partial Y^{lmn}}}+X{\pa\over{\pa g}}\right)
\gamma^i{}_j,
&\betasoft d\cr
}$$
where we define $V^{(ij)}=V^{ij}+V^{ji}$, etc., and where 
\eqn\gam{
(\gamma_1)^i{}_j={\cal O}\gamma^i{}_j,}
\eqn\calO{
{\cal O} = \left(Mg^2{\partial\over{\partial g^2}}-h^{lmn}{\partial
\over{\partial Y^{lmn}}}\right)}
and
\eqn\ty{
\tY^{ijk}= Y^{l(jk}(m^2)^{i)}{}_l.}

A word on scheme dependence. The results given in Eq.~\betasoft{}\ are 
valid in the $\DREDp$ (or $\drbar$) scheme, introduced in 
Ref.~\ref\jjmvy{I.~Jack et al, \prd50 (1994) R5481} and established 
to all orders in Ref.~\ahetal. They remain valid under a class of 
linked redefinitions of $g$ and $M$\jj\jjpb, and in particular in a scheme 
where the NSVZ form of the gauge $\beta$-function is valid.

\newsec{The conformal anomaly solution}

In a previous paper\jjpa\ we showed that on 
an RG trajectory $Y^{ijk}(g)$, the
following relations are RG-invariant:
\eqna\concc$$\eqalignno{h^{ijk}  &= -M g\frakk{dY^{ijk}}{dg}, &\concc a\cr
b^{ij} &= -Mg\frakk{d\mu^{ij}}{dg}, &\concc b\cr
(m^2)^i{}_j &= \frakk{g^2}{2\beta_g}|M|^2\frakk{d\ga^i{}_j}{dg}. &\concc c\cr
}$$
Remarkably, these relations can be supplemented by the following equation, 
which is also RG invariant:
\eqn\gaugino{M=M_0{\beta_g\over g},}
where $M_0$ is a constant. Presently we will verify the RG invariance of 
this equation; let us accept this for the moment, 
and substitute in Eq.~\concc{}. Noting that
\eqn\betay{\beta_Y^{ijk}=\beta_g\frakk{dY^{ijk}}{dg},\qquad
\beta_{\mu}^{ij}=\beta_g\frakk{d\mu^{ij}}{dg},}
we obtain
\eqna\result$$\eqalignno{
h^{ijk}=&-M_0\beta_Y^{ijk},&\result a\cr
b^{ij}=&-M_0\beta_{\mu}^{ij}, &\result b\cr
(m^2)^i{}_j =& \frak{1}{2}|M_0|^2\mu\frakk{d\ga^i{}_j}{d\mu}.&\result c\cr}$$
This is the form of the anomaly-generated relations as presented and 
applied in Ref.~\con, except that $M_0$ becomes the gravitino mass, 
$m_{\frakk{3}{2}}$. Several of the papers in Ref.~\con\ remark that 
Eqs.~\gaugino, \result{}\ are in fact RG invariant; 
our purpose here is to establish this result 
to all orders in a general theory. 
It will be clear from our ``bottom-up'' approach that this is the unique 
RG-invariant integral of the equations that can be written 
without knowing the explicit form of the $\beta$-functions. 

In Ref.~\jjpa\ we demonstrated the RG invariance of Eqs.~\concc{};
we made, however,  certain assumptions  (which are
spelt out later). In fact, these assumptions were not necessary, 
and it is possible to show that Eqs.~\gaugino\ and
\result{}\ are RG-invariant  for a completely  general (non-finite)
theory. Technical details are postponed to Appendix~A, but we give the broad
outlines here. 
The essential point is that the nature of the solution makes ${\cal O}$
a total derivative when acting on the relevant structures. 
If we substitute Eqs.~\gaugino\ and \result{a}\ in Eq.~\calO, we can 
show using Eq.~(A.5) that 
\eqn\calOa{
{\cal O}\beta_g=\frakk12M_0\mu{d\over{d\mu}}\beta_g,}
and hence using Eq.~\betasoft{a}\ 
we find immediately 
\eqn\RGinv{
\beta_M=M_0\mu{d\over{d\mu}}\left({\beta_g\over g}\right),}
which shows that Eq.~\gaugino\ is RG-invariant 
for $M_0$ constant. (The RG invariance 
of this equation was actually demonstrated for a theory with no Yukawa 
couplings in Ref.~\ref\his{J.~Hisano and M.~Shifman, 
\prd56 (1997) 5475}.)   

We can also show that (see Eq.~(A.7))  
\eqn\calOc{
{\cal O}\gamma^i{}_j=\frakk12M_0\mu{d\over{d\mu}}\gamma^i{}_j,}
and hence, using Eqs.~\gam, \result{a}\ in Eq.~\betasoft{b}, we obtain
\eqn\betaha{
\beta_h^{ijk}=-M_0\left[\gamma^{(i}{}_l\beta_Y^{jk)l}
+\mu{d\gamma^{(i}{}_l\over{d\mu}}Y^{jk)l}\right]
=-M_0\mu{d\over{d\mu}}[\gamma^{(i}{}_lY^{jk)l}] = - M_0\mu{d\over{d\mu}}
\beta_Y^{ijk},}
demonstrating that
Eq.~\result{a}\ is RG-invariant. The RG invariance of Eq.~\result{b}\
follows similarly. Finally, 
we can show that upon using Eqs.~\result{}\ in Eq.~\betasoft{d}, we just obtain
\eqn\betama{
(\beta_{m^2})^i{}_j=
\frakk12|M_0|^2\left(\mu {d\over{d\mu}}\right)^2\gamma^i{}_j,}
which confirms the RG invariance of Eq.~\result{c}.

The authors of Ref.~\akk\ derived results 
similar to our Eqs.~\betasoft{},
the major difference being that they omit the $X$-term in Eq.~\betasoft{d}, 
apparently believing that it would be absent given a \sic\ regulator.
Here we disagree; consider, for instance, a general NSVZ scheme. 
By this we mean one for which
$\beta_g$
is given by\ref\shifetal{V.~Novikov et al,
\npb 229 (1983) 381\semi
V.~Novikov et al, \plb166 (1986) 329\semi
M.~Shifman and A.~Vainstein, \npb 277 (1986) 456} 
\eqn\shifa{
\beta_g^{\NSVZ}={g^3\over{16\pi^2}}
\left[{Q-2r^{-1}\tr[\gamma C(R)]\over{
1-2g^2C(G)(\lf)^{-1}}}\right],}
for which we showed\jjpb\ 
that $X$ is given by
\eqn\shifb{
X^{\NSVZ}=-2{g^3\over{16\pi^2}}
{S\over{\left[1-2g^2C(G)(16\pi^2)^{-1}\right]}},}
where 
\eqn\defs{
Q = T(R) - 3C(G)\qquad S =  r^{-1}\tr [m^2C(R)] -MM^* C(G),}
and  $r$ is the number of generators of the gauge group and
$C(R)$ and $C(G)$ are the quadratic matter and adjoint Casimirs respectively.
$X$ as given in Eq.~\shifb\ is clearly non-zero, and remains so in  
perturbatively related schemes such as $\drbar$; indeed, $X$ 
plays a crucial role in establishing the RG invariance of Eq.~\result{c}
(see Eq.~(A.10)). However, it is nevertheless possible, in special cases, 
to find a scheme 
in which $X$ vanishes. Consider a theory with
$\gamma^i{}_j$ and $(m^2)^i{}_j$ diagonal and define a transformation from the
NSVZ scheme to a new ``holomorphic'' scheme by
\eqn\hol{
{1\over{\alpha^{{\rm H}}}}={1\over{\alpha^{\NSVZ}}}+\frakk{2}{\lf}
C(G)\ln\alpha^{{\NSVZ}}
-2{1\over{\lf}}[F(Y)+F(Y^*)])}
(where $\alpha=g^2$), such that
\eqn\fdefa{
\sum_aR_{ai}{\pa F\over{\pa Y_a}}Y_a=T(R_i),}
where $Y_a$ are the independent Yukawa couplings and $R_{ai}$ counts the 
number of fields $\Phi^i$ at the vertex $Y_a$.
(This scheme, which was introduced in Ref.~\jjpb, 
is related but not identical to the 
usual  holomorphic scheme, for a discussion of which see
Refs.~\ref\ahma{N.~Arkani-Hamed and H.~Murayama, \prd57 (1998) 6638, 
hep-th/9707133}.)
It is then easy to check using Eq.~\shifa, \shifb, together with the 
transformation rule for $X$ given in Ref.~\jjpb, that $\beta_{\alpha^H}$ is
one-loop exact and $X^H$ vanishes to all orders. As an example, if there
at least as many independent Yukawa couplings than there are fields, 
then we may in general define
a matrix $\tilde R_{ia}$ such that $\tilde R_{ia}R_{aj}=\delta_{ij}$. We may 
then take $F$ in Eq.~\fdefa\ to be given by
\eqn\fdefb{
F(Y_a)=\sum_iT(R_i)\tilde R_{ia}\ln(Y_a).} 

We end our discussion of the conformal anomaly case with a remark on the
 generalisation to a gauge group with a direct product structure 
($G_1\otimes G_2\cdots$). In fact, Eq.~\result{}\ is unaffected, but we 
must replace Eq.~\gaugino\ by  
\eqn\gauginob{M_I=M_0{\beta_{g_I}\over g_I}.} 
Thus at a unification scale $M_X$ such that $g_I = g_X$ for all
$I$,  the gaugino masses do not themselves unify. Their ratios  are, 
however, determined by the  values of the corresponding gauge $\beta$-functions
at $M_X$, that is, by the ratios of the one-loop $\beta$-function 
coefficients (in the one-loop approximation). We note also that if one or
more  of the gauge group factors is abelian, then the Fayet-Iliopoulos
$D$-term is renormalised, even if set zero at (say) $M_X$.  If $D$ is
eliminated using its equation of motion then  this renormalisation gives
rise to additional contributions to $\beta_{m^2}$  beyond those included
in Eq.~\betasoft{d}; see for example Eq.~(11) and Eq.~(16)  of 
Ref.~\ref\jja{I.~Jack and D.R.T.~Jones, \plb 333 (1994) 372}. Our results 
do, however, apply to the abelian case as long as we do not so eliminate 
$D$ {\it ab initio}. For a discussion of the phenomenological consequences 
of the linear $D$-term in the MSSM, see Ref.~\ref\gfm{ A. de Gouv\^ea,
A. Friedland and H. Murayama \prd 59 (1999) 095008}. 

\newsec{The finite case}

We turn now to the finite case, corresponding  to a choice of 
RG trajectories 
$Y^{ijk}(g)$ such that $\beta_g(g,Y(g))=\gamma(g,Y(g))=0$. Since in the NSVZ 
scheme $\gamma = 0$ implies $\beta_g = 0$ if $Q = 0$, it is clear 
that there is a large class of $N = 1$ theories of this type.  
Now in this case, Eqs.~\gaugino, \result{}\ give
zero for all the soft parameters. However, we know that there are non-trivial 
values for soft couplings consistent with finiteness, at least through  
two loops\jja; and we can  
in fact generalise this to all orders by demonstrating 
the existence of a distinct set of RG-invariant trajectories for this case. 
We begin by replacing Eq.~\gaugino\ with 
\eqn\gauginof{M = M_0,}
where $M_0$ is constant. Now from Eq.~\betasoft{a,b}\ it is clear that 
to get $\beta_M = \beta_h = \beta_b = 0$ we must have 
\eqn\mhf{
{\cal O}\left({\beta_g\over g}\right) = {\cal O}\gamma = 0.} 
The solution to this is evidently
\eqna\conccff$$\eqalignno{h^{ijk}  & = -M_0 g\frakk{dY^{ijk}}{dg}
 &\conccff a\cr
b^{ij} &= \hbox{const} &\conccff b\cr}$$
(compare with Eq.~\concc{a,b}.)
This was shown in Ref.~\ref\kaz{D.I.~Kazakov, \plb421 (1998) 211},
at least in the case of real couplings. The case of general couplings 
requires some additional assumptions, however. Once again 
denoting the independent Yukawa couplings by $Y_a$, we need the following 
identities:  
\eqna\conce$$\eqalignno{
Y_a\frakk{\pa\ga_i}{\pa Y_a}
= &Y_a^*\frakk{\pa\ga_i}{\pa Y_a^*},
 \quad(\hbox{no sum on $a$})&\conce{a}\cr
Y'_a\frakk{\pa\ga_i}{\pa Y_a}
= &Y^{\prime*}_a\frakk{\pa\ga_i}{\pa Y_a^*},
 \quad(\hbox{no sum on $a$}).&\conce{b}\cr}$$
Notice that in the non-finite case, Eq.\conce{b}\ follows from 
Eq.~(A.7); but in the finite case Eq.~(A.7) is trivial and we 
have been unable to establish either of Eq.~\conce{a,b} for general 
couplings, though they are clearly true, for example, if $Y_a (g)$ can 
be chosen to be real.
It follows from Eqs.~\conce{b}\ and \conccff{a}\ that
\eqn\concfff{
{\cal O}\equiv \frakk{1}{2}M_0 g\frakk{d}{dg},}
and hence we see from 
Eqs.~\betasoft{a-c}\ that $\beta_M=\beta_h=\beta_b=0$. 

Let us now turn to the case of $m^2$. It is clear that we must re-examine the
derivation of Eq.~\concc{c} in Ref.~\jjpa, since it is ill-defined in the 
finite case. This derivation proceeded by 
proposing a relation for $m^2$ of the form
\eqn\newi{m^2_i = \frakk{1}{3}|M_0|^2 f_i(g),}
where we assume that $(m^2)^i{}_j$ is diagonal and that 
$(m^2)^i{}_j=m^2_i\delta^i{}_j$. (In view of Eq.~\betasoft{d}, for consistency 
we must also assume $\gamma^i{}_j=\gamma_i\delta^i{}_j$.)
We now need to ensure that when 
Eqs.~\conccff{a}\ and Eq.~\newi\ are substituted into Eq.~\betasoft{d}, 
we obtain $\beta_{m_i^2}=0$.
To prove this, we need an identity which follows from
Eqs.~\conce.
These equations yield immediately
\eqn\ratio{
{Y'_a\over{Y_a}}={Y_a^{\prime *}\over{Y_a^*}}.}
Acting on Eq.~\conce{a}\ with ${Y'_aY'_b\over{Y_a}}
{\pa\over{\pa Y_b}}$, and using Eq.~\ratio, we obtain 
\eqn\ident{
Y_a^{\prime *}Y_b'{\pa^2\ga_i\over{\pa Y_a^*\pa Y_b}}=Y_a'Y_b'
{\pa^2\ga_i\over{\pa Y_a\pa Y_b}}+{Y_a^{\prime2}\over{Y_a}}
{\pa\ga_i\over{\pa Y_a}}.}
Using this identity, together with Eq.~\conce{b}, we can show (assuming that
on the RG trajectory $X=|M_0|^2\Xtilde(g)$) that
\eqn\newm{\beta_{m_i^2} = |M_0|^2\left[\frakk{1}{2}g^2\ga_i'' +
\left(\frakk{3}{2}g + \Xtilde\right)\ga_i'\right]=0}
provided we choose 
\eqn\fdef{
\frakk13R_{ai}f_i=\left(\frakk{3}{2}g + \Xtilde\right)\frakk{Y'_a}{Y_a}
+\frakk{1}{2}g^2\left[\frakk{Y_a''}{Y_a}- \left(\frakk{Y_a'}{Y_a}\right)^2
\right],\quad(\hbox{no sum on $a$})}
where $R_{ai}$ was defined earlier.

In our previous work we assumed $f_i (g) = f(g)$ for all $i$; this is 
sufficient but not necessary for an RG invariant 
solution\ref\kkm{Y.~Kawamura, T.~Kobayashi and J.~Kubo, \plb405 (1997) 64\semi
T.~Kobayashi et al, \npb511 (1998) 45\semi
T.~Kobayashi, J.~Kubo, and G.~Zoupanos, 
\plb 427 (1998) 291\semi
Y.~Kawamura, T.~Kobayashi and J.~Kubo, \plb 432 (1998) 108
}.
The challenge is now to solve Eq.~\fdef, a task which is complicated by
the fact that $\Xtilde$ depends on $f_i$, as can be seen from the explicit
expression in the NSVZ scheme, Eq.~\shifb. 

In the finite case, the results of Ref.~\akk\ were used (again omitting 
$X$) in Ref.~\kaz\ to derive a result equivalent to Eq.~\fdef.
However, even in the finite case there is no reason to expect $X$ to vanish in 
general.
For example, consider the case of a finite trajectory for a single real Yukawa 
coupling $Y$ and a single chiral field. Suppose moreover that we are
working in an NSVZ scheme as described above. The finite trajectory is given 
by $Y=ag+bg^5$ for some constants $a$, $b$
(there is no $O(g^3)$ term
since one-loop finiteness automatically implies two-loop finiteness\ref\fin{
D.R.T.~Jones and L.~Mezincescu, \plb136 (1984) 242\semi
P.~West, \plb136 (1984) 371\semi
A.~Parkes and P.~West, \plb138 (1984) 99\semi
D.R.T.~Jones and L.~Mezincescu, \plb138 (1984) 293}).
Then one can in principle solve Eqs.~\fdef\ and \shifb, and the result for 
$X$ is
clearly non-zero at $O(g^7)$.

It is interesting to compare the leading order 
results for the scalar masses in the general (Eq.~\result{c}) 
and finite (Eq.~\fdef) cases. Neglecting Yukawa couplings, Eq.~\result{c}\
becomes 
\eqn\slep{
(m^2)^i{}_j = - \frakk{2g^4}{(\lf)^2} |M_0|^2 Q C(R)^i_j}
 so that $m^2 < 0$ for a non-asymptotically free gauge group. 
This causes problems with slepton masses in the MSSM; 
various ways round this are discussed in Ref.\con. 
On the other hand, Eq.~\fdef\ gives 
\eqn\mtr{\sum_i R_{ai}m_i^2 = |M_0|^2.}
Finite GUT phenomenology attracts occasional interest
\ref\jrab{D.R.T. Jones and S. Raby, \plb 143 (1984) 137\semi
J. E. Bjorkman,  D.R.T. Jones and S. Raby, \npb 259 (1985) 503\semi
J. Leon et al  \plb 156 (1985) 66\semi 
A.V. Ermushev, D.I. Kazakov and  O.V. Tarasov \npb 281 (1987) 72\semi
D. Kapetanakis, M. Mondragon and G. Zoupanos, \zpc 60 (1993) 181\semi
N. G. Deshpande, Xiao-Gang He and E. Keith, \plb 332 (1994) 88\semi 
D.I. Kazakov et al \npb 471 (1996) 389\semi
J. Kubo and M. Mondragon and G. Zoupanos,  
Nucl.Phys.Proc.Suppl.~56{\bf B} (1997) 281}
but is lacking a convincing derivation from 
an underlying theory. In the non-finite case, on the other hand, 
the conformal anomaly justification is very elegant\con, and 
is being pursued energetically in spite of the difficulty 
mentioned above. Unfortunately, all solutions proposed so far 
sully to some extent the purity of the conception. 

In Appendix~A we present some technical details associated with 
our derivation, and in Appendix~B some remarks on the Method of 
Characteristics as applied to the $\beta$-function equations. 
\appendix{A}{}
In Ref.~\jjpa, we demonstrated the RG invariance of Eqs.~\concc{}\ in some 
detail for a theory with one real Yukawa coupling and a single superfield, 
and we also explained how to extend our results to the general case, under
the special assumptions that $\gamma$ is diagonal and that Eq.~\conce{a}\ holds.
However, we can show that Eqs.~\result{} are RG-invariant for a completely 
general theory with arbitrary complex Yukawa couplings. The general strategy 
is as followed in Ref.~\jjpa\ and sketched out above; if Eqs.~\result{}\ are
applied, we can write the $\beta$-functions as given in Eqs.~\betasoft{}\ 
in terms of total $\mu$-derivatives. However, in the general case we 
need a number of identities which are stated and proved here.   

We start with the demonstration of RG invariance of Eq.~\gaugino. 
We define 
\eqn\defq{
\Qcal=\sum_{klm}\beta_Y^{klm}{\pa\over{\pa Y^{klm}}}, \qquad \Qcal^*
=\sum_{klm}\beta_{Yklm}{\pa\over{\pa Y_{klm}}}.}
Now on use of Eq.~\result{a}, we find 
\eqn\calOb{
\Ocal=\frakk12M_0\left(\beta_g{\pa\over{\pa g}}+2\Qcal\right),}
while 
\eqn\muderiv{
\mu{d\over{d\mu}}=\beta_g{\pa\over{\pa g}}+\Rcal,}
where
\eqn\calQ{
\Rcal=\Qcal+\Qcal^*.}
However, it is easy to show that
\eqn\deriv{\Qcal^*\beta_g=\Qcal\beta_g, }
as follows: If we think in terms of the superfield diagrams contributing to 
$\beta_g$,
then in view of Eq.~\betay, the effect of the operator on the left/right hand 
side of Eq.~\deriv\ is to 
give the sum of insertions of $\gamma$ onto each line emanating from each 
$Y$/$Y^*$ vertex, respectively. Each line emanating from a $Y$ vertex ends at 
$Y^*$ vertex, possibly after passing through some 
matter/gauge vertices. Since
$\gamma$ commutes with the gauge generators, the effect of each operator is the 
same, namely to give the sum of insertions of a $\gamma$ onto each line joining 
a $Y$ and a $Y^*$ vertex. In view of Eq.~\deriv, on comparing 
Eqs.~\calOb, \muderiv\ we verify Eq.~\calOa, enabling
us to obtain Eq.~\RGinv.

Turning now to $h^{ijk}$, the crucial step in the derivation of Eq.~\betaha\
was the use of Eq.~\calOc. 
It is easy to see, using the same argument as for Eq.~\deriv, that
for any quantity $X^i{}_j$ constructed according to supergraph Feynman rules,
we have
\eqn\master{
(\Qcal X)^i{}_j-(\Qcal^* X)^i{}_j=[\gamma,X]^i{}_j.}
Eq.~\deriv\ follows immediately, as does
\eqn\mgamma{\Qcal\gamma = \Qcal^*\gamma}
whence Eq.~\calOc\ follows, leading immediately to Eq.~\betaha.

To show the RG invariance of Eq.~\result{c}, however, 
requires some care. The equation we wish to prove, Eq.~\betama, translates using
Eq.~\muderiv\ to
\eqn\betamb{
(\beta_{m^2})^i{}_j=\frakk12|M_0|^2\left(\beta_g{\pa\over{\pa g}}
+\Rcal\right)^2\gamma^i{}_j.}
Two identities are required to prove this starting from Eq.~\betasoft{d}. 
The first (which follows by repeated application of Eq.~\master) is
\eqn\rsq{
\Rcal^2\gamma^i{}_j=4\beta_Y^{klm}\beta_{Ypqr}
{\pa^2\over{\pa Y^{klm}\pa Y_{pqr}}}
\gamma^i{}_j+(\Rcal\gamma)^{(k}{}_nY^{lm)n}{\pa\gamma^i{}_j\over{\pa Y^{klm}}}
+(\Rcal\gamma)^n{}_{(k}Y_{lm)n}{\pa\gamma^i{}_j\over{\pa Y_{klm}}}.}
The second identity is that if Eqs.~\result{a,b}\ are imposed, then  
\eqn\Xident{
|M_0|^2\mu{d\beta_g\over{d\mu}}=3{\beta_g^2|M_0|^2\over{g}}+2X.}
An equivalent identity was deduced in Ref.~\jjpa\ (Eq.~(30), with $A=0$); 
we can also
prove it directly using the explicit forms for $\beta_g$ and $X$ in the
NSVZ scheme given in Eqs.~\shifa, \shifb. 
Finally, expanding out $\Ocal\Ocal^*$ in Eq.~\betasoft{d},
applying Eq.~\result{}, and comparing with the result of expanding out the 
right-hand side of Eq.~(A.8)\ and using Eqs.~\rsq, \Xident, we find that 
Eq.~\betama\ is satisfied.

\appendix{B}{}

Here we indicate the relationship between 
the conformal anomaly solution given in 
Eqs.~\gaugino, \result{}\ and the 
{\it characteristics\/}  
of Eqs.~\betasoft{}. For simplicity we restrict 
ourselves to the case of one real Yukawa coupling. Eqs.~\betasoft{a,b}\ may be 
written as
\eqna\betasoftn$$\eqalignno{
\beta_{M'}=&\left(M'{\pa\over{\pa g}}-h{\pa\over{\pa Y}}\right)\beta_g,
&\betasoftn a\cr
\beta_h=&-\left(M'{\pa\over{\pa g}}-h{\pa\over{\pa Y}}\right)\beta_Y,
&\betasoftn b\cr}$$
where $M'=gM$.
Beginning with Eq.~\betasoftn{a}, the gradients of the characteristics 
are given by 
\eqn\chara{
\frakk{1}{M_0}dt = \frakk{dg}{M'} = -\frakk{dY}{h} 
= \frakk{1}{\beta_{M'}}d\beta_g}
while Eq.~\betasoftn{b} gives
\eqn\charb{
\frakk{1}{M_0}dt = \frakk{dg}{M'} = -\frakk{dY}{h} 
= -\frakk{1}{\beta_h}d\beta_Y.}
Together these yield:
\eqna\charc$$\eqalignno{M_0\frakk{dg}{dt} & = M'& \charc a\cr
M_0\frakk{dY}{dt} & = -h & \charc b\cr
M_0\frakk{d\beta_g}{dt}& = \beta_{M'}& 
\charc c\cr
M_0\frakk{d\beta_Y}{dt}& = -\beta_h & \charc d\cr 
}$$
It can then be seen that requiring RG invariance of 
Eqs.~\charc{a,b} leads to Eqs.~\charc{c,d}\ 
and therefore Eqs.~\charc{a,b}\ furnish RG-invariant
expressions for $M'$ and $h$. Indeed if we identify $t$ with $\mu$, then we
regain Eqs.~\gaugino\ and \result{a}. It is clear that Eq.~\result{b}\ will
follow similarly. We have not pursued this approach with 
Eq.~\betasoft{d}\ except to note that (in the real case) 
this second order PDE is parabolic, 
with characteristic gradient $\frakk{dY}{dg} = -h/M'$, in common with 
Eqs.~\betasoftn{}\ above. 

\bigskip\centerline{{\bf Acknowledgements}}\nobreak

This work was supported in part by a Research Fellowship from the 
Leverhulme Trust. We thank Hugh Osborn for pointing out a couple of 
typographical errors in the original version. 

\listrefs
\bye